\documentclass[conference]{IEEEtran}
\IEEEoverridecommandlockouts
\usepackage{booktabs}
\usepackage{cite}
\usepackage{amsmath,amssymb,amsfonts}
\usepackage{algorithmic}
\usepackage{graphicx}
\usepackage{textcomp}
\usepackage{xcolor}
\usepackage[T1]{fontenc}
\usepackage{hyperref}

\makeatletter
\newcommand{\linebreakand}{%
  \end{@IEEEauthorhalign}
  \hfill\mbox{}\par
  \mbox{}\hfill\begin{@IEEEauthorhalign}
}
\makeatother

\def\BibTeX{{\rm B\kern-.05em{\sc i\kern-.025em b}\kern-.08em
    T\kern-.1667em\lower.7ex\hbox{E}\kern-.125emX}}
\begin{document}



\title{\textsf{sbom-unifier}: Integration Framework for Heterogeneous SBOMs
}

\author{\IEEEauthorblockN{Yusuke Moriwaki}
\IEEEauthorblockA{\textit{Ritsumeikan University}\\
Osaka, Japan \\
is0746ps@ed.ritsumei.ac.jp}
\and
\IEEEauthorblockN{Tetsuya Kanda}
\IEEEauthorblockA{\textit{Notre Dame Seishin University}\\
Okayama, Japan \\
kanda@m.ndsu.ac.jp}
\and
\IEEEauthorblockN{Yuki Manabe}
\IEEEauthorblockA{\textit{The University of Fukuchiyama}\\
Kyoto, Japan \\
manabe-yuki@fukuchiyama.ac.jp}
\linebreakand
\IEEEauthorblockN{Shi Qiu}
\IEEEauthorblockA{\textit{Toshiba Corporation}\\
Kanagawa, Japan \\
shi.qiu.b79@mail.toshiba
}
\and

\IEEEauthorblockN{Shiyu Yang, Erina Makihara, Norihiro Yoshida, and Katsuro Inoue}
\IEEEauthorblockA{\textit{Ritsumeikan University}\\
Osaka, Japan \\
{(yangsy, makihara, norihiro, inoue-k)@fc.ritsumei.ac.jp}
}}



\maketitle

\begin{abstract}
A Software Bill of Materials (SBOM) is a machine-readable inventory of software components, increasingly required for vulnerability management and license compliance. 
%
%
However, existing SBOM generation tools often leave many  SPDX-defined fields missing or only partially populated, because different tools produce heterogeneous outputs with uneven field-level coverage.
We present \textsf{sbom-unifier}, a framework that improves SBOM completeness through field-level integration and complementation of multiple tool outputs and file-level enrichment. Unlike existing tools that simply concatenate SBOMs without identifying records referring to the same component, \textsf{sbom-unifier} identifies components via Package URL (PURL), complements missing field values by a deterministic priority-based strategy, reconstructs cross-section references, and further enriches file-level fields. 

Across 90 open-source projects in 9 programming languages, \textsf{sbom-unifier} preserves high completeness for required fields and, over the 39 SPDX 2.3 required and optional fields, raises the fully covered rate by 8 percentage points and reduces the totally missing rate by 11 percentage points over the respective best-performing 
individual tools.
\textsf{sbom-unifier} is available at 
\url{https://github.com/MoriwakiYusuke/sbom-unifier}.
Also, a short video is available at
\url{https://youtu.be/QgjcEADFB1w}.

\end{abstract}

\begin{IEEEkeywords}
software bill of materials, SPDX, software supply chain, field-level merge
\end{IEEEkeywords}

\section{Introduction}

Modern software development increasingly depends on a large ecosystem of third-party components, particularly open-source software (OSS). While this trend accelerates development, it also introduces challenges in managing security risks and license compliance~\cite{synopsys2025}.

A key mechanism for addressing these challenges is the Software Bill of Materials (SBOM), which provides a machine-readable representation of software components and their dependencies~\cite{ntia_sbom}. The Software Package Data Exchange (SPDX), standardized by the Linux Foundation, defines a structured format of fields for representing SBOMs~\cite{spdx_official}.


SBOMs represent software as sets of components described by metadata such as name, version, license, dependency, and others. However, existing SBOM generation tools often provide incomplete field availability due to their design policy, resulting in missing or partially missing fields~\cite{Wang_2026}.

To address this limitation, we propose \textsf{sbom-unifier}, a framework that improves SBOM completeness through complementation-based merging and file-level enrichment using external tools. The system identifies components across input SBOMs with PURL (Package URL) as a common identifier, complements missing field values from peer records, and enriches missing fields through lightweight file inspection.
Furthermore, \textsf{sbom-unifier} includes an evaluation framework that quantitatively measures field-level completeness, enabling direct comparison with individual tool outputs.
\textsf{sbom-unifier} is intended for users who need to integrate, inspect, and compare SBOMs generated by multiple tools for security and compliance workflows.






The contributions of this paper are as follows:

\begin{itemize}
\item We design and implement \textsf{sbom-unifier}, a framework that integrates heterogeneous SBOMs generated by multiple tools and complements missing metadata through file-level enrichment.

\item Through an empirical evaluation on 90 open-source projects, we show that \textsf{sbom-unifier} improves field-level completeness compared with individual SBOM generation tools.
\end{itemize}

\section{Background and Related Work}

\subsection{SBOM Generation Tools and Completeness}

SBOM generation tools produce machine-processable SBOMs from target software artifacts. Depending on the software lifecycle stage, SBOMs are categorized into types such as design, source, build, and runtime SBOMs~\cite{cisa_sbom_types}. Among these, source SBOMs are widely used in practice, as they provide a fundamental view of software composition~\cite{landscape_study}. In this paper, we focus on source SBOMs.

Various SBOM generation tools have been developed, including Microsoft SBOM Tool\footnote{{\texttt{https://github.com/microsoft/sbom-tool}}}, Syft\footnote{{\texttt{https://github.com/anchore/syft}}}, Trivy\footnote{{\texttt{https://github.com/aquasecurity/trivy}}}, and GitHub Dependency Graph\footnote{\texttt{https://docs.github.com/\allowbreak en/\allowbreak code-\allowbreak security/\allowbreak supply-\allowbreak chain-\allowbreak security/\allowbreak understanding-\allowbreak your-\allowbreak software-\allowbreak supply-\allowbreak chain/\allowbreak about-\allowbreak the-\allowbreak dependency-\allowbreak graph}}. These tools are designed with different analysis goals, such as dependency extraction, vulnerability scanning, or repository metadata analysis~\cite{manzi2025sbom}. As a result, the set of fields in the SBOMs they generate varies significantly~\cite{accuracy_sbom,correctness_sbom}, and no single tool achieves comprehensive metadata completeness of SPDX-defined fields~\cite{Wang_2026}.

Even for required fields, tools may leave some fields blank, and coverage becomes substantially lower when optional fields are included. This incompleteness of SBOM metadata can negatively affect downstream tasks such as vulnerability detection and license compliance analysis~\cite{impacts_sbom_vuln}. Section~VI quantitatively evaluates this limitation and shows how \textsf{sbom-unifier} improves field-level completeness.

\subsection{SBOM Merging}

Existing approaches to improving SBOM completeness include merging multiple SBOMs. We distinguish two strategies based on how they treat entries that describe the same package or file.

\textbf{Concatenation-based merging.} Most existing tools, such as CycloneDX CLI\footnote{{\texttt{https://github.com/CycloneDX/cyclonedx-cli}}}, SPDXMerge\footnote{{\texttt{https://github.com/philips-software/SPDXMerge}}}, and sbommerge\footnote{{\texttt{https://github.com/anthonyharrison/sbommerge}}}, concatenate the entries of input SBOMs section by section (Packages, Files, Relationships, etc.) without checking whether two entries describe the same package or file. Consequently, the same component appears as duplicate entries, missing fields in one entry are not filled in from another entry of the same component, and cross-section references (e.g., a Relationship pointing to a Package by its SPDX ID) still point to the pre-merge entries instead of the merged ones.

\textbf{Complementation-based merging.} An alternative strategy is to identify, across input SBOMs, entries that describe the same package or file, and to complement missing or \texttt{NOASSERTION} values (which indicate that no claim is made in SPDX) in one entry using the corresponding entry in another. FatBOM\footnote{\href{https://github.com/VexStore/fatbom}{\texttt{https://github.com/VexStore/fatbom}}}
invokes multiple SBOM generators and merges results based on package identification by name and version; however, it neither performs per-field complementation---filling missing field values from peer records---nor preserves original cross-section references such as Relationships, instead regenerating them uniformly. Furthermore, FatBOM does not support PURL-based identification, which limits its ability to align packages across tools with heterogeneous naming conventions. To our knowledge, no existing tool combines PURL-based entry identification, per-field complementation, and reconstruction of cross-section references into a single workflow.


These limitations motivate our approach, which we describe in the following section.


\begin{figure*}[t]
\centering
\includegraphics[width=0.9\textwidth]{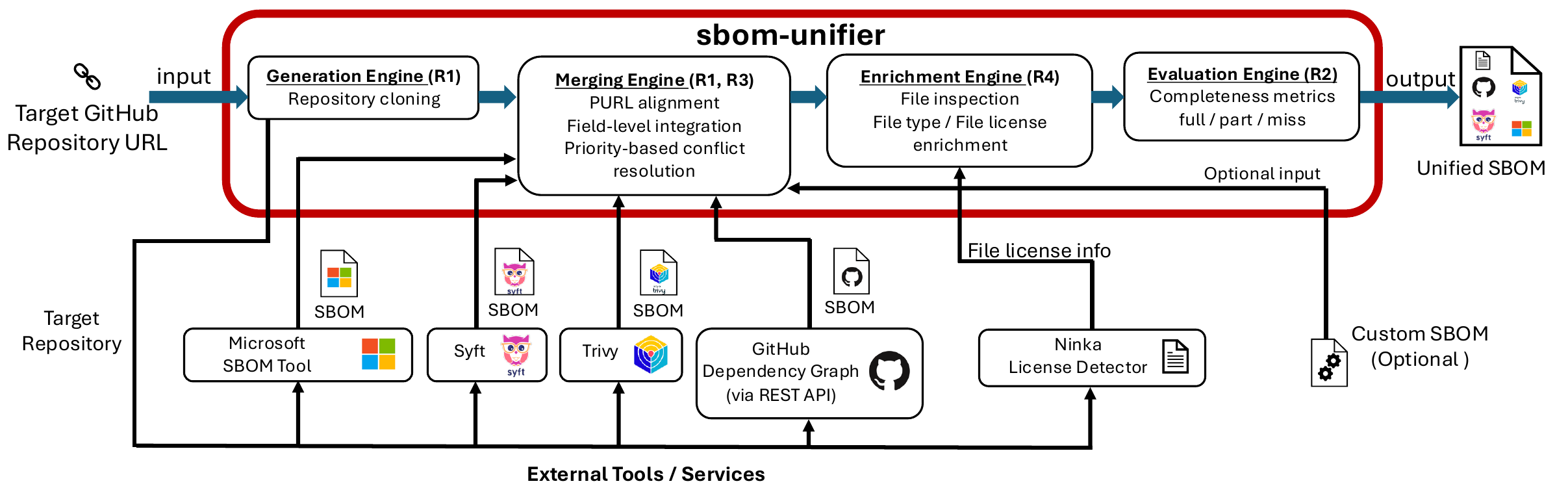}
\caption{System overview of \textsf{sbom-unifier}.}
\label{fig:architecture}
\end{figure*}

\section{SBOM Fields and Evaluation Scope}
\label{sec:SBOMFields}

An SBOM defined in SPDX consists of multiple fields that describe software components and their associated metadata. 
We assign each field to one of the following attributes based on the SPDX specification, with additional criteria for Omitted fields.

\begin{itemize}
  \item \textbf{Required}: Fields that must be present.
  \item \textbf{Optional}: Fields that may be included but not mandatory.
  \item \textbf{Deprecated}: Fields that are no longer recommended for use.
  \item \textbf{Omitted}: Fields that are not commonly used in practice and are therefore excluded from the evaluation.
\end{itemize}

The \textit{deprecated} fields correspond to those explicitly marked as deprecated in SPDX 2.3 specification. The \textit{omitted} fields include (i) optional SBOM sections that are not typically generated by existing tools (e.g., Snippet Information, Other Licensing Information, and Annotation Information), and (ii) fields that require manual input or external curation, such as comment and attribution-related fields.

In this study, we evaluate only the fields categorized as \textit{required} or \textit{optional}. This results in a total of 39 evaluated fields, consisting of 14 required fields and 25 optional fields. The complete list of evaluated fields, grouped by SPDX section and annotated with their requirement status, is presented in Table~\ref{tab:spdx-fields}.

\begin{table}[!t]
\centering
\caption{Evaluated SPDX 2.3 fields by section and requirement status (R: required, O: optional). The 39 evaluated fields comprise 14 required and 25 optional fields (deprecated and omitted fields are excluded).}
\label{tab:spdx-fields}
\scriptsize
\setlength{\tabcolsep}{6pt}
\renewcommand{\arraystretch}{.6}
\begin{tabular}{@{}l l c@{}}
\toprule
\textbf{Section} & \textbf{Field} & \textbf{R/O} \\
\midrule
Document Creation & SPDX Version                & R \\
                  & Data License                & R \\
                  & SPDX Identifier             & R \\
                  & Document Name               & R \\
                  & SPDX Document Namespace     & R \\
                  & License List Version        & O \\
                  & Creator                     & R \\
                  & Created                     & R \\
\midrule
Package           & Package Name                & R \\
                  & Package SPDX Identifier     & R \\
                  & Package Version             & O \\
                  & Package File Name           & O \\
                  & Package Supplier            & O \\
                  & Package Originator          & O \\
                  & Package Download Location   & R \\
                  & Files Analyzed              & O \\
                  & Package Checksum            & O \\
                  & Package Home Page           & O \\
                  & Source Information          & O \\
                  & Concluded License           & O \\
                  & Declared License            & O \\
                  & Copyright Text              & O \\
                  & External Reference          & O \\
                  & Package Attribution Text    & O \\
                  & Primary Package Purpose     & O \\
                  & Release Date                & O \\
                  & Built Date                  & O \\
                  & Valid Until Date            & O \\
\midrule
File              & File Name                   & R \\
                  & File SPDX Identifier        & R \\
                  & File Type                   & O \\
                  & File Checksum               & R \\
                  & Concluded License           & O \\
                  & License Information in File & O \\
                  & Copyright Text              & O \\
                  & File Contributor            & O \\
                  & File Attribution Text       & O \\
\midrule
Relationship      & Relationship                & R \\
                  & Relationship Comment        & O \\
\bottomrule
\end{tabular}
\end{table}

Based on these definitions, we consider two evaluation scopes:

\begin{itemize}
  \item \textbf{Required Scope}: fields categorized as required
  \item \textbf{Required+Optional Scope}: fields categorized as required or optional
\end{itemize}

These definitions are used consistently in the evaluation described in Section~\ref{sec:eval-results}.


\section{Design Principles and Requirements}

\textsf{sbom-unifier} is designed to support configurable and reproducible SBOM integration workflows targeting the fields defined in Section~\ref{sec:SBOMFields}.
The framework is based on four principles: component alignment, field complementarity, deterministic resolution, and measurable completeness.
Based on these principles, \textsf{sbom-unifier} satisfies the following requirements.

\textbf{R1: Complementation-based merging across  heterogeneous SBOMs.}
The framework identifies components across SBOMs using a common identifier (PURL) and complements missing field values from peer records, in contrast to concatenation-based merging that performs only a structural union without record identification.

\textbf{R2: Inspection and comparison of field-level completeness.} 
The framework measures field-level completeness and enables comparison between individual tools and the unified SBOM.

\textbf{R3: Reproducible and configurable conflict handling.} 
The framework resolves conflicting field values using a deterministic priority strategy.

\textbf{R4: SBOM enrichment via lightweight file inspection.} 
The framework enriches missing fields through lightweight file inspection and external tools.


\section{Architecture of \textsf{sbom-unifier}}
\label{sec:arch}


\begin{figure}[t]
\centering
\includegraphics[width=0.80\linewidth]{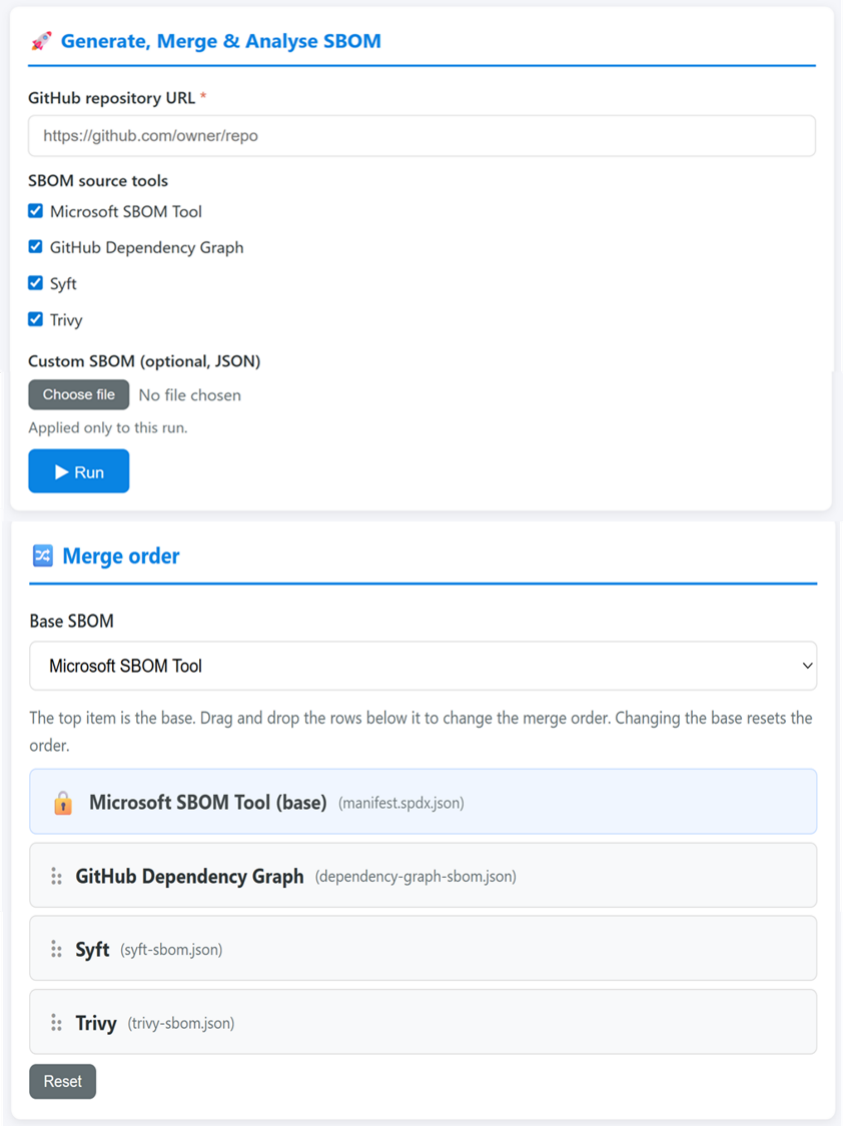}
\caption{Input interface for configuring \textsf{sbom-unifier}.}
\label{fig:demo1}
\end{figure}

\begin{figure}[t]
\centering
\includegraphics[width=0.80\linewidth]{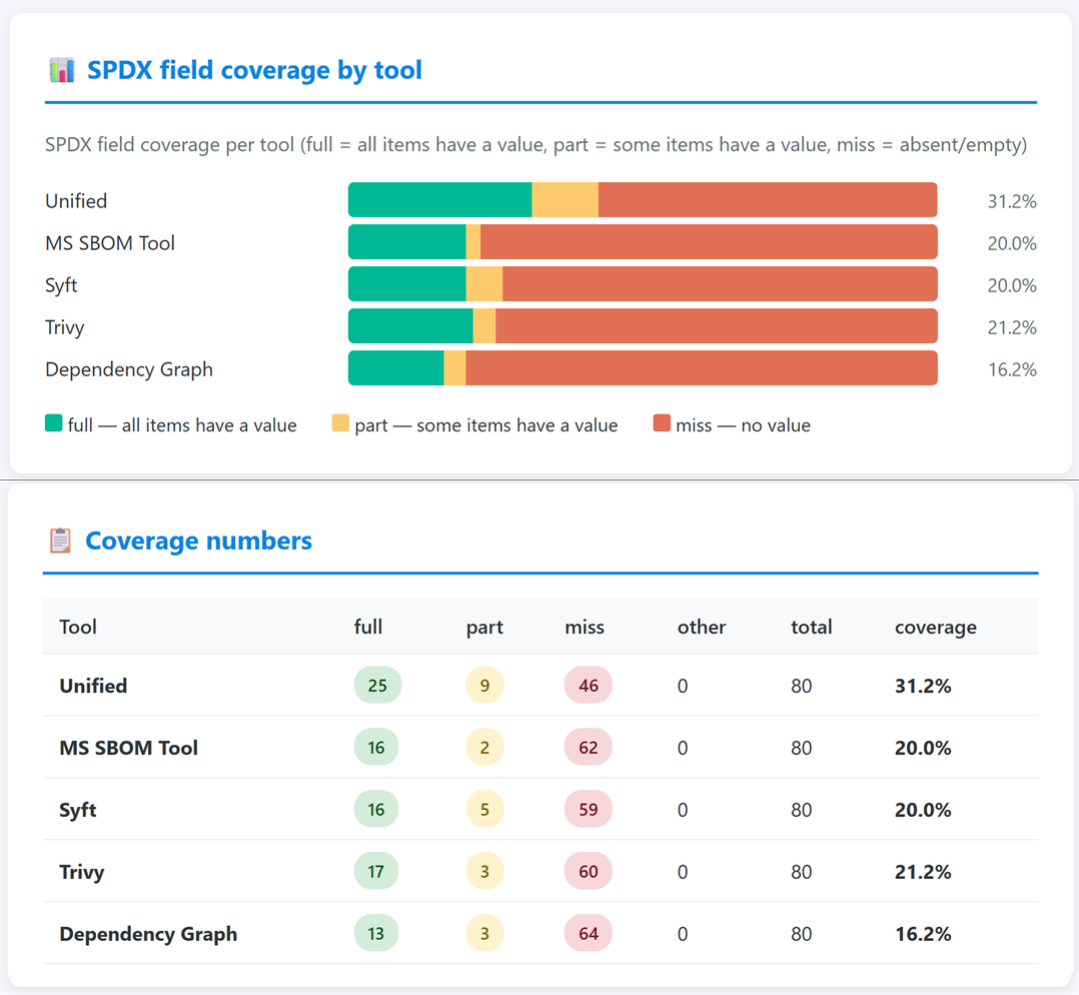}
\caption{Result interface showing field-level completeness.}
\label{fig:demo2}
\end{figure}

\textsf{sbom-unifier} takes a target repository URL and generates a unified SBOM together with a field-level completeness evaluation. As shown in Fig.~\ref{fig:architecture}, the framework consists of four components corresponding to R1--R4. A typical workflow starts by selecting a target repository, SBOM generation tools, and merge settings, and ends with the generation of a unified SBOM together with a field-level completeness summary.

The Generation Engine (R1) collects SBOMs from external tools, including Microsoft SBOM Tool, Syft, Trivy, and GitHub Dependency Graph, while also accepting externally generated SBOMs as optional inputs. The input configuration is specified through the interface shown in Fig.~\ref{fig:demo1}, where users select target repositories, choose SBOM generation tools, and configure merge priorities.

The Merging Engine (R1, R3) performs complementation-based merging by aligning packages using Package URL (PURL) and files using file paths. Missing or \texttt{NOASSERTION} values are complemented using peer SBOM records describing the same component. PURL was selected because it provides ecosystem-independent package identification across heterogeneous SBOM generators. Entries without a PURL are excluded from this alignment, and unmatched entries are retained as distinct packages.

The Enrichment Engine (R4) complements fields that cannot be recovered through cross-tool complementation alone. In particular, it enriches file-related metadata such as file types and license information using lightweight file inspection and the external tool Ninka~\cite{ninka}.

The Evaluation Engine (R2) summarizes field-level completeness, enabling users to assess the effects of integration and enrichment. The resulting completeness statistics and unified SBOM are presented through the interface shown in Fig.~\ref{fig:demo2}.

\section{Evaluation}

\subsection{Evaluation Setup}

The objective of this evaluation is to quantitatively assess how \textsf{sbom-unifier} improves field-level completeness of source SBOMs. Throughout this evaluation, we report \emph{field-level coverage}, defined as the proportion of evaluated SPDX fields populated within their corresponding records, classified into full, part, and miss categories (described in \ref{sec:evaluation_criteria}). To ensure a fair comparison, we use the same set of SBOM generation tools (Microsoft SBOM Tool, GitHub Dependency Graph, Syft, and Trivy) and compare their individual outputs with the SBOM produced by \textsf{sbom-unifier}, which merges and enriches these outputs.

We conduct the evaluation following the architecture described in Section~\ref{sec:arch}. The process is as follows:

\begin{enumerate}
\item A target repository is cloned from GitHub.
\item \textsf{sbom-unifier} is executed, which internally invokes multiple SBOM generation tools, merges their outputs by complementation, and enriches missing fields.
\item Field-level coverage (described also in \ref{sec:evaluation_criteria}) is computed for both the individual tool outputs and the unified SBOM, and the results are compared under identical conditions.
\end{enumerate}

For reproducibility, the merge order is fixed during the evaluation. The output of Microsoft SBOM Tool is used as the base SBOM, and Dependency Graph, Syft, and Trivy are applied as sources in this order, followed by the output of the Enrichment Engine.

The target repositories consist of 90 public GitHub projects, with 10 highly starred ones selected from each of 9 programming languages (C\#, Go, Java, JavaScript, PHP, Python, Ruby, Rust, and Swift)  
and at least 95\% language purity.  
We evaluate two scopes, Required Scope and Required+Optional Scope.

\subsection{Evaluation Criteria}
\label{sec:evaluation_criteria}

Each field is classified into one of three categories:

\begin{itemize}
\item full: generated for all target projects
\item part: generated for at least one but not all target projects
\item miss: not generated for any target project
\end{itemize}


The coverage is computed as the proportion of these categories over all evaluated fields.

\subsection{Evaluation Results}
\label{sec:eval-results}

\begin{figure}[t]
\centering
\includegraphics[width=\linewidth]{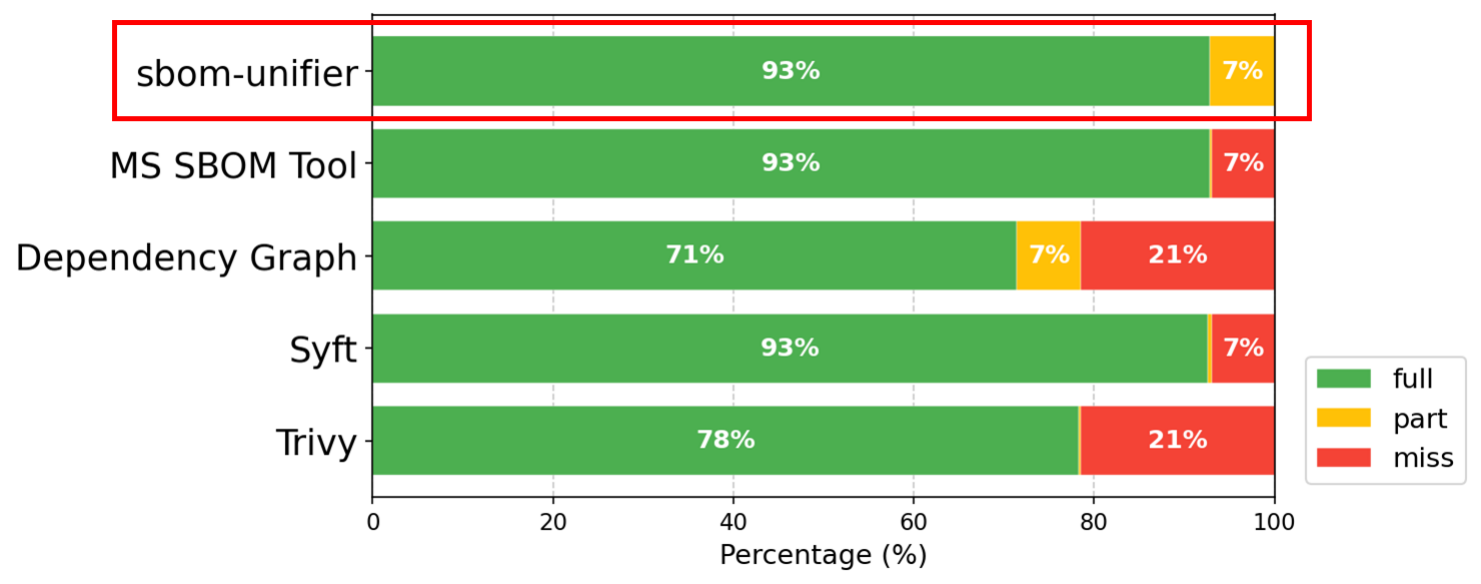}
\caption{Field-level coverage in the Required Scope (14 fields), averaged over 90 projects across 9 programming languages (10 projects per language). The proposed method is highlighted in red.}
\label{fig:required}
\end{figure}

\begin{figure}[t]
\centering
\includegraphics[width=\linewidth]{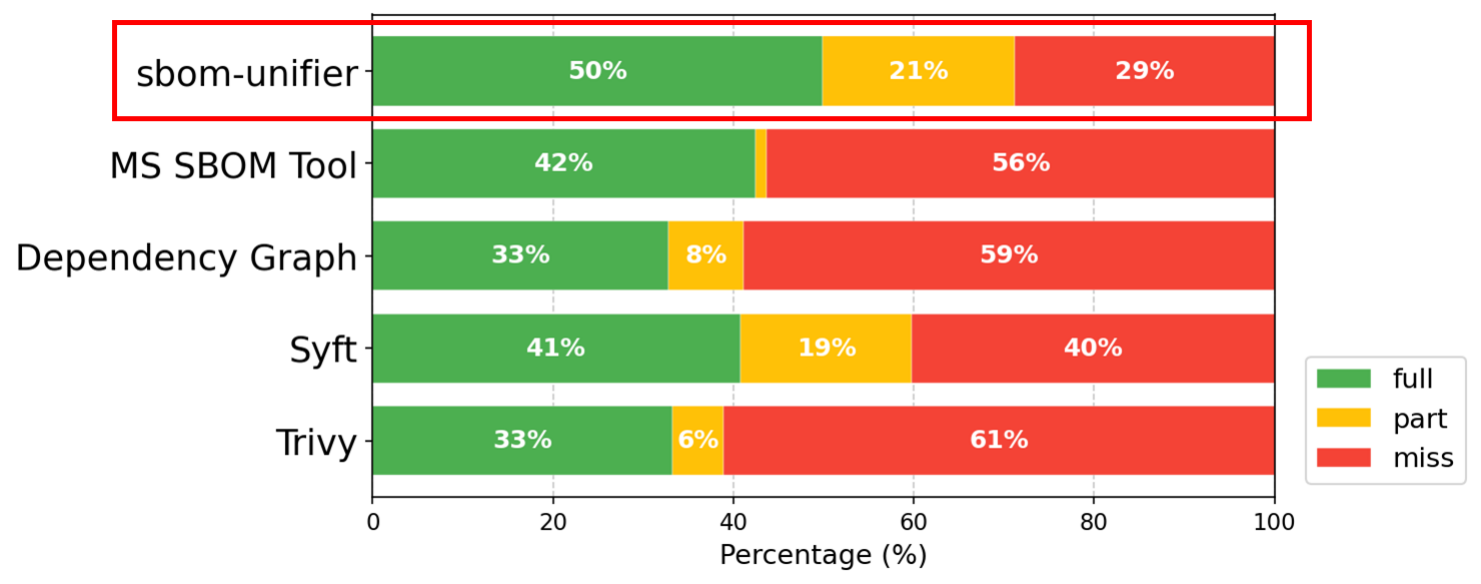}
\caption{Field-level coverage in the Required+Optional Scope (39 fields), averaged over 90 projects across 9 programming languages (10 projects per language). The proposed method is highlighted in red.}
\label{fig:required_optional}
\end{figure}

Fig.~\ref{fig:required} and Fig.~\ref{fig:required_optional} present the field-level coverage for the Required Scope and the Required+Optional Scope, respectively.
In both figures, full indicates that all projects populated the field, part indicates some, and miss indicates none.

In the Required Scope, the upper bound on field-level completeness is already attained by the strongest individual tools: \textsf{sbom-unifier}, Microsoft SBOM Tool, and Syft all reach 93\% full, while Dependency Graph and Trivy retain a 21\% miss rate. 
Although the full rate remains identical to the strongest individual tools, \textsf{sbom-unifier} further reduces missing fields by complementing them as partially populated fields (turning red into orange in Fig.~\ref{fig:required}).

In the Required+Optional Scope, \textsf{sbom-unifier} attains 50\% full and 29\% miss, 
outperforming the best individual tools (42\% full with Microsoft SBOM Tool; 40\% miss with Syft), 
while Trivy and Dependency Graph leave 59--61\% of fields missing.
These results indicate that complementation-based merging and file-level enrichment
recover information that no single tool can provide.
This result is consistent across the nine languages, with per-language full rates ranging from 48.7\% to 51.3\% (standard deviation: 0.8 percentage points).

\subsection{Discussion}


The two scopes expose complementary aspects of \textsf{sbom-unifier}'s behavior.
In the Required Scope, \textsf{sbom-unifier} preserves the completeness level of the strongest individual tools without degradation.
In the Required+Optional Scope, the improvement over the best individual tool (50\% vs.\ 42\% full) confirms that the gains reflect genuine complementation across heterogeneous tools and file-level enrichment, not mere redistribution of existing information.

Some fields remain missing because they essentially require external metadata or manual curation. Addressing these fields would require integration with additional data sources such as package registries or curated datasets.

\subsection{Limitations}

The correctness of the generated SBOMs largely depends on the outputs of the underlying SBOM generation tools. While \textsf{sbom-unifier} improves field-level completeness,
it does not verify or correct the values provided by the underlying tools.
Since an incorrect value can be more harmful than a missing one for downstream tasks, assessing and improving the correctness of field values is an important future direction of this work.

The generated SBOM quality could be further improved by incorporating deeper analysis of the target software, such as source-code inspection, or by leveraging external information sources (e.g., web-based metadata). However, such mechanisms are beyond the scope of the current implementation.

Another limitation is that the current implementation supports only SPDX 2.3 because, at the time of writing, it is the only format common to all of the integrated generation tools. Although the proposed approach is not specific to this format, extending the system to support newer standards such as SPDX 3.0 and other formats like CycloneDX remains future work.






\section{Conclusion}

We presented \textsf{sbom-unifier}, a tool that improves the completeness of source SBOMs through complementation-based merging and file-level enrichment. 
Unlike concatenation-based merging, \textsf{sbom-unifier} identifies components 
by PURL, complements missing or \texttt{NOASSERTION} field values from peer records, 
and reconstructs cross-section references to preserve document consistency.



Through an empirical evaluation on 90 open-source projects, we showed that \textsf{sbom-unifier} improves field-level completeness compared to individual SBOM generation tools. In particular, the system preserves high completeness for required fields while reducing missing values across optional fields.



This work suggests that improving SBOM quality should be treated not merely as document aggregation, but as a structured software metadata integration problem.
The current architecture separates alignment and conflict-handling policies, enabling future integration of more advanced conflict-resolution strategies, including semantic or AI-based approaches.

\label{sec:availability}
\textbf{Artifact Availability:}
\textsf{sbom-unifier} is published as open source under the MIT License at \url{https://github.com/MoriwakiYusuke/sbom-unifier}, associated with the replication package at
\url{https://github.com/MoriwakiYusuke/sbom-unifier-replication}. 
A pre-built image is distributed on Docker~Hub as\\ 
\texttt{yusukemoriwaki\allowbreak/sbom-unifier\allowbreak:latest}.

\textbf{Acknowledgment:}

This work is partially supported by JSPS JP23K28065, JP24K02923, JP24K14895, and JST JPMJCR25U7. We thank Yusaku Kato and the anonymous reviewers for their valuable feedback. Claude Code assisted parts of the implementation, and Claude and ChatGPT were used for English translation and proofreading. All content was validated by the authors.

\bibliographystyle{IEEEtran}
\bibliography{bib/references}


\end{document}